\documentclass[prl,twocolumn,superscriptaddress]{revtex4}

\usepackage{graphicx}

\begin{document}

\title{Spin wave theory for antiferromagnetic XXZ spin model on a triangle lattice in the presence of an external magnetic field}
\author{J. Y. Gan$^1$, F. C. Zhang$^2$, Z. B. Su}
\affiliation{Institute of Theoretical Physics, Chinese Academy of Sciences, Beijing, People's Republic of China\\
$^2$ Department of Physics, University of Cincinnati, Cincinnati OH 45221, USA
}

\begin{abstract}
Spin wave theory is applied to a quantum antiferromagnetic XXZ model on a triangle lattice in the presence of an in-plane
magnetic field. The effect of the field is found to enhance the
quantum fluctuation and to reduce the sublattice magnetization
at the intermediate field strength in the anisotropic case.
The possible implication to the field driven quantum
phase transition from a spin solid to a spin liquid is discussed.

\end{abstract}


\maketitle

\section*{\bf 1. Introduction}
Quantum spin systems in low dimensions continue to be an interesting subject. 
Quantum fluctuation generally destroys the long range order in one dimension\cite{bethe},
and plays an important role in two dimension~\cite{anderson}. In this paper, we study
the ground state of spin-$S$ antiferromagnetic quantum XXZ model on a triangle
lattice in the presence of an external in-plane magnetic field. 

Our main focus is on the effect of the magnetic field to the quantum fluctuation.
It is well known that a spin system becomes polarized at a very high field (above
a threshold), which suppresses the quantum fluctuation. It is, however, less
clear if the magnetic field enhances or suppresses the quantum fluctuation
at intermediate field strength. The latter problem may be important to systems
near the boundary of a quantum solid (ordered phase) to quantum liquid (disorder phase)
transition.  
Our interest on this problem is partially motivated by the recent experiments on $Cs_2CuCl_4$.
That system~\cite{coldea1} is a quasi 2-dimensional $S=1/2$ 
frustrated Heisenberg antiferromagnet with a
weak anisotropy favoring spins aligned in the basal plane, likely due to the 
Dzyaloshinskii-
Moriya interaction~\cite{d-m}. Coldea et al.~\cite{coldea2,coldea3} 
have used neutron scattering to study the ground state and the dynamics of the system
in high magnetic field. Among the observations, these authors found 
that the system undergoes phase transitions from a spin solid to a spin liquid
to a spin fully polarized states as the in-plane magnetic field increases.
Their experiment has raised an interesting theoretical question of the
effect of the field on the quantum fluctuation and on the possible field driven
quantum spin liquid. The basal plane of $Cs_2CuCl_4$ consists of a triangle lattice with
the spin-spin coupling much stronger along one direction than along other two 
directions. There have been some theoretical works attempting to address
some of the issues relevant to the experiments~\cite{bocquet,chung,shen}. Here
we shall consider a simplified model to study the field effect on the quantum
spin fluctuation. For simplicity, in our model we shall neglect 
the lattice anisotropy in the real systems, and assume the spin couplings are the same
between  all the pairs of the nearest neighboring sites. We shall use 
XXZ model to describe the anisotropy in spin space, which distinguishes the
in-plane field from the perpendicular field.

We apply spin-wave theory~\cite{mattis} to study the leading order correction to 
the sublattice magnetization due to the quantum fluctuation.    
The spin wave theory or equivalent theories 
have been applied to a number of systems,
including the
antiferromagnetic Heisenberg model in a triangle lattice~\cite{yoshioka} and
the quantum XY model~\cite{mit}. We calculate the ground state energy and
the sublattice magnetization as functions of the magnetic field, and find that
the quantum fluctuation measured by the reduction of the sublattice magnetization
is enhanced as the field increases at the intermediate field strength 
and in the anisotropic case. The possible implication to the field driven quantum
phase transition will be discussed.

This paper is organized as follows. In section 2, we introduce the model Hamiltonian.
Section 3 describes the application of the spin wave theory to the present model.
The results for the ground state energy and the sublattice magnetization are presented in section 4.
In the final section, we summarize our results and discuss their possible implications
to the field driven quantum spin liquid.

\bigskip
\section*{\bf 2. Model Hamiltonian}
We consider spin-$S$ antiferromagnetic XXZ model on a triangular lattice
of $3N$ sites in the x-z plane
in the presence of an external magnetic field $h$ along the z-axis. The
lattice constant is set to be 1.
The Hamiltonian can be written as follows:
\begin{eqnarray}
H = \sum\limits_{i,\delta}(S_i^x S_{i+\delta}^x + S_i^z
S_{i+\delta}^z + w S_i^y  S_{i+\delta}^y)
- S h \sum\limits_{i} S_i^z . \label{hamiltonian}
\end{eqnarray}
In the above equation, $S_i^{\mu}$ is the $\mu$ component of the spin-S
operator on site $i$, and
$\delta$ denotes the three nearest-neighbor bond vectors (see Fig.1)
given by, $\delta_{\alpha}=(1,0)$, $\delta_{\beta}=(- 1/2, \sqrt{3}/2)$,
and $\delta_{\gamma} = (-1/2,-\sqrt{3}/2)$.
$hS$ is the strength of the magnetic field, and $h$ will be considered to be order
of unity for the purpose of analyzing the
spin wave theory, so that the magnetic field contribution to the energy
is of the same order as the leading order spin-spin interaction energy.
$w$ is a parameter describing the anisotropy of the model,
and $0 \leq w \leq 1$.
The model is reduced to the Heisenberg model at $w=1$, and to the XY model
at $w=0$. Note that the mdoel with $w > 1$ belongs to a different universal
class and will not be discussed here.

\begin{figure}
\includegraphics*[width=3.in]{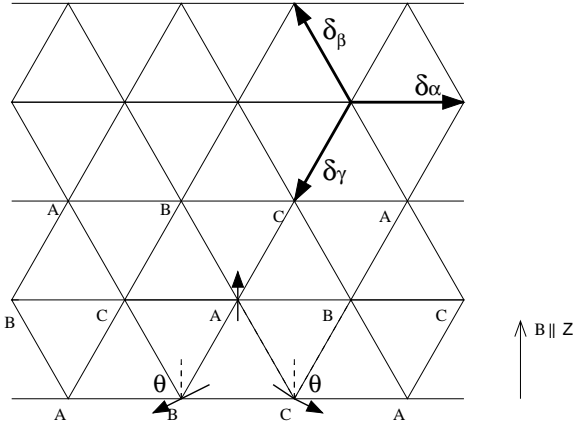}
\caption{Three sublattices in a triangle lattice and the classical
spin state studied in this paper.}
\label{fig:pd}
\end{figure}

\bigskip
\section*{\bf 3. Spin-wave theory}
We apply the spin wave theory to study this model~\cite{mattis}. This requires a
proper choice of the classical state
upon which the spin wave represents minimum quantum fluctuations.
Since $w \leq 1$, we choose
quantization axes within the $x-z$ plane, and consider the spin
quantizations in three sublattices
as indicated in Fig. 1. Namely, the spins in sublattice $A$ are aligned
along the z-axis and
the spins in sublattices $B$ and $C$ are aligned along the $\theta$ and
$ -\theta$ directions respectively, where $\theta$ is determined
variationally. Such a quantization is consistent with the expected three sublattice
classical ground state.

We introduce three boson annihilation operators $a, b, c$ and their
corresponding creation operators
by means of the Holstein-Primakoff transformation~\cite{holstein} to represent spins on
sublattices $A$, $B$, and $C$, respectively.
They are given by $S_i^{z'} = S - n_i$, $S_i^{+} = \sqrt{2S - n_i}\,d_i$,
and $S_i^{-} = d_i^{\dagger} \sqrt{2S - n_i}$, where $d =a, b, c$ for the site
$i \in A, B, C $
respectively.
$n_i=d_i^{\dagger}d_i$, and $z'$ is the corresponding spin
quantization axis, and $S_i^{\pm} = S_i^{x'} \pm iS_i^{y'}$.
We apply the standard spin wave theory and write the Hamiltonian
in terms of large $S$-expansion.
The leading order term is proportional to $S^2$, which is the energy of the classical
ground state given by
\begin{eqnarray}
E_{cl}= NS^2[6\cos{\theta} + 3\cos {2\theta} -h(1+2\cos{\theta})].
\end{eqnarray}
The angle $\theta$ is determined variationally by $dE_{cl}/d\theta =0$, from which we find
\begin{eqnarray}
\cos{\theta} = \left \{  \begin{array}{cc} (h-3)/6,  \, \, h \leq h_c = 9\\
                  \,  \,    \\
               1, \,  \, h \geq h_c
               \end{array}
               \right.
\end{eqnarray}
Note that the value of $\theta$ is independent of the spin anisotropic
parameter $w$ for the $y$-component
of the spin does not play any role in the classical limit.
At zero field, we obtain $\theta =120^o$, recovering the well known result.
The case of $\theta =0$ ($h \geq h_c$) corresponds to the
ferromagnetic phase where all the spins are
fully polarized along the field direction. We see that as
the field increases, $\theta$ decreases from
$120^o$ to $0$, and the classical antiferromagnetic ground
state is gradually transformed into a ferromagnetic state.

The leading order quantum fluctuation to the classical solution is proportional to
$S$, and it is given by,
\begin{eqnarray}
H_{FL}=3S\sum\limits_{\vec k}D_{\vec k}^{\dagger} {\cal M}_{\vec k} D_{\vec k} + E^0_{FL}(h), \label{h_fl}
\end{eqnarray}
where
\begin{eqnarray}
E^0_{FL}(h) =   \left \{  \begin{array}{cc}
- 9 NS/2, \,\, h \leq h_c\\
 \\
-3NS(h/2-3), \, \, h \geq h_c
        \end{array}
        \right.
\end{eqnarray}
In Eq. (4),  the sum is over the reduced Brillouin zone, and
$D^{\dagger}_{\vec k}= ( A^{\dagger}_{\vec k}, \tilde{A}_{-\vec k})$,
$ A^{\dagger}_{\vec k} = (a^{\dagger}_{\vec k},  b^{\dagger}_{\vec k}, c^{\dagger}_{\vec k})$,
and $\tilde{A}$ is the transpose of $A$.
$\cal M$ is a 6 $\times$ 6 matrix, and can be written
in terms of an idendity matrix $\cal I$ and the Pauli matrix $\sigma_x$ as
\begin{eqnarray}
{\cal M}= {\cal I} \otimes( M_0 + M_{+}) + \sigma_x \otimes M_{-}
\end{eqnarray}
where  $M_0$ is a 3 $\times$ 3 diagonal matrix, whose matrix elements are given by
$ (M_0)_{11}= \frac{h}{6} -\cos{\theta}$, and
  $(M_0)_{22} = (M_0)_{33} = \frac{1}{6}(h\cos{\theta} - 3\cos{\theta} - 3\cos{2\theta})$, and
\begin{eqnarray}
 M_{\pm} = \left ( \begin{array}{ccc}
 0   & (\cos{\theta} \pm w)\gamma_{\vec k} & (\cos\theta \pm w)\gamma_{\vec k}^{*}   \\
 (\cos \theta \pm w)\gamma_{\vec k}^{*}  & 0 & (\cos 2\theta \pm w)\gamma_{\vec k}       \\
 (\cos \theta \pm w)\gamma_{\vec k} & (\cos 2\theta \pm w)\gamma_{\vec k}^{*}   & 0
        \end{array}
 \right )
\end{eqnarray}
In the above equations, $\gamma_{\vec k}= (1/12) \sum_{\delta}
\exp{(i\vec{k}\cdot\vec{\delta})}$, where the sum of $\delta$ runs over
$\{\delta_\alpha,\delta_\beta,\delta_\gamma \}$. The off diagonal matrices
$M_{\pm}$ represents the interaction among the three sublattices. The
$\theta$ term in the matrix elements arises from the $x$- and $z$- components
of the spin fluctuation and the $w$ term arises from the $y$- component
of the spin fluctuation.

$H_{FL}$ can be diagonalized by using the Bogoliubov transformation. We
introduce three
bosonic operators $\alpha_{n,\vec k}$ with $n = 1, 2, 3$, and vector operators
${\cal A}^{\dagger}_{\vec k}= (\alpha^{\dagger}_{1,\vec k},\alpha^{\dagger}_{2,\vec k}, \alpha^{\dagger}_{3,\vec k})$.
and ${\cal D}^{\dagger}_{\vec k} = ( {\cal A}^{\dagger}_{\vec k}, \tilde{\cal A}_{- \vec k} )$.
They are related to the original Holstein-Primikoff
boson operators,
$D_{\vec k}$,  by a Bogoliubov transformation $U$, and that
 $H_{FL}$ is diagonalized in terms of these new boson
operators,
\begin{eqnarray}
D_{\vec k} = U {\cal D}_{\vec k}\\
H_{FL} = 2S\sum\limits_{n,\vec k} \omega_{n,\vec k}
(\alpha^{\dagger}_{n,\vec k} \alpha_{n,\vec k} + 1/2 ) + E^0_{FL}(h)
\end{eqnarray}
where $\omega_{n,\vec k}$ (n = 1, 2, 3) are the energy dispersions of the three
spin wave excitations (magnon modes)
in the spin ordered ground state.

In general, the Bogoliubov transformation can be carried out  hence the magnon
dispersions can be calculated
numerically. This will be discussed in the following section.
In the ferromagnetic phase, $h \geq h_c$,  we have analytical results,
\begin{eqnarray}
\omega_{n, \vec k} = \sqrt{[-3 + \frac{h}{2} + 3(1+ w)T_{n,\vec k}]^2 -
[3(1-w)T_{n,\vec k}]^2}       \nonumber
\end{eqnarray}
where
$T_{n,\vec k}=\gamma_{\vec k} \xi^n + \gamma_{\vec k}^{*} \xi^{2n}$, with
$\xi=\exp{(i2\pi/3)}$. Note that $ -1/4 \leq T_{n, \vec k} \leq 1/2$.
This result
is in agreement with the previous work~\cite{guillou}.

\bigskip
\section*{\bf 4. Results}
In this section, we present the results for the ground state energy and the sublattice magnetization
of the antiferromagnetic XXZ model obtained from the spin wave theory described in the previous section.

\bigskip
\subsection*{\bf 4.1. Ground state energy}
In this subsection, we discuss the ground state energy up to the
leading order quantum fluctuation, namely
to the order of $S$ in the large $S$-expansions.
The ground state energy per bond  (there are total $9N$ bonds in the lattice) is given by
\begin{eqnarray}
\epsilon_0 = \epsilon_{cl} + \epsilon_{fl}
\end{eqnarray}
where $\epsilon_{cl}$ and $\epsilon_{fl}$ are the classical and the fluctuation energies, respectively.
From Eq. (2), $\epsilon_{cl}$ is given by,
\begin{eqnarray}
\epsilon_{cl} & = &\left \{ \begin{array}{ll} -(\frac{1}{2} + \frac{h^2}{54})S^2, & h \leq h_C \\
                                 (1 - \frac{h}{3})S^2,  & h\geq h_C
                  \end{array}
                  \right.
\end{eqnarray}
To obtain $\epsilon_{fl}$, we include the zero point energy of the bosons
in Eq. (9), which leads to
\begin{eqnarray}
\epsilon_{fl} & = &\left \{ \begin{array}{ll}
(-\frac{1}{2} +  \frac{1}{9N} \sum \limits_{n,\vec k}\omega_{n,\vec k})S, & h \leq h_C \\
 (- (\frac{h}{6} - 1) + \frac{1}{9N} \sum \limits_{n,\vec k}\omega_{n,\vec k})S,  & h\geq h_C
                  \end{array}
                  \right.
\end{eqnarray}

\begin{figure}
\includegraphics*[width=3.in,angle=270]{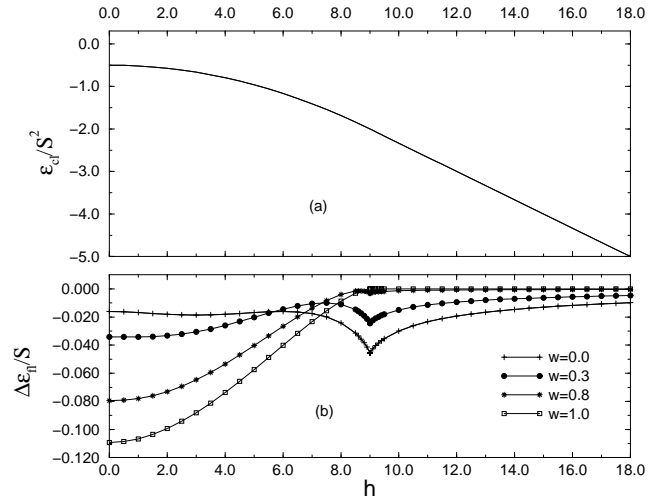}
\caption{(a) The classical energy per bond as function of magnetic field $h$
         (b) The leading order quantum correction to the energy in spin wave theory
as functions of $h$ for several values of $w$.  }
\label{fig:pd}
\end{figure}

We have solved the Bogoliubov transfromation problem numerically
to diagonalize $H_{FL}$ and to calculate the dispersions
$\omega_{n, \vec k}$ and the energy $\epsilon_0$.
In Fig. 2(a), we plot the energy of the classical ground state
as a function of the magnetic field $h$. The energy decreases monotonically
as $h$ increases, and is linear in $h$ in the
ferromagnetic state with
$h \geq h_C$. In Fig. 2(b), we show the quantum correction to the ground state energy
 as functions of $h$ for several values of $w$.
Firstly, we note that   the quantum correction to the energy is very small in comparison
with the classical energy even for $S=1/2$. This indicates that the spin wave theory
is a good approximation for the present model.
At $h=0$, the quantum correction is largest in the Heisenberg model ($w=1$) and
smallest in the XY model ($w=0$). This is consistent with the
intuition that the $y-$ component of the spin  would increase the quantum fluctuation.
 As $h$ increases, however, the magnitude of the quantum fluctuation
as a function of $w$ reverses the order as we can see from Fig. 2(b).
This becomes apparent at $h = h_C$, where
the quantum correction to the energy in the Heisenberg model vanishes, while that in
the XY model reaches the maximum.
It is interesting to note that the magnetic field enhances the quantum fluctuation
in  certain parameter region of the mdoel, a point we will come back for further discussion in the
next subsection. It is also interesting to note that the leading order quantum
fluctuation remains finite in the general case even in the ferromagnetic phase.
The Heisenberg limit of the model is only a
special case where the leading order
quantum fluctuation vanishes. The latter
can be understood by examining the matrices in $H_{FL}$. At $h \geq h_C$, and $w=1$,
we have $M_- =0$, so that boson annihilation operators
$(a, b, c)$ do not mix with their creation operators $(a^{\dagger}, b^{\dagger}, c^{\dagger})$,
hence the system is similar to the ferromagnetic Heisenberg model and has
zero quantum fluctuation.

\bigskip
\section*{\bf 4.2. Sublattice magnetization}
We now move to our main interest to discuss the quantum corrections to the sublattice magnetization.
The magnetization in sublattice $L =\{A, B, C \}$ is defined
as the average spin component within the same sublattice along its
quantization axis,
\begin{eqnarray}
\langle S^{z'}_L \rangle  = \frac{1}{N}\sum_{i \in L} \langle S^{z'}_i \rangle
                                 =  S - \langle \Delta S_L \rangle
\end{eqnarray}
where $S$ is the classical value of the spin, and the second term,
$\Delta S_L = (1/N) \sum\limits_{i \in L} \langle d^{\dagger}_i d_i \rangle$,
is the reduction from the classical spin value due to the quantum fluctuation, and
$\langle Q \rangle$ is the expectation value of operator $Q$ in the ground state.
 $\Delta S_L$ can be calculated from the Bogoliubov transformation matrix $U$,
\begin{eqnarray}
\Delta S_L & = & \frac{1}{N}\sum \limits_{\vec k}|U_{l4}|^2 + |U_{l5}|^2
+ |U_{l6}|^2
\end{eqnarray}
where $l=\{1,2,3\}$ for $L=\{A,B,C\}$ respectively, and $U$ is obtained in the diagonalization
of $H_{FL}$.  The results are shown in Fig. 3 for $\Delta S_A$ and in Fig. 4 for $\Delta S_B$
for functions of the field $h$ for several values of $w$.
By symmetry, $\Delta S_C = \Delta S_B$.

We first discuss the Heisenberg limit of $w=1$. At $h =0$, we find $\Delta S_L \approx 0.26$
for all the three sublattices. This result is the same as that reported early using a Schwinger
boson mean field theory by Yoshioka~\cite{yoshioka}. As $h$ increases, both $\Delta S_A$ and
$\Delta S_B$ decrease, but $\Delta S_A$ drops much faster as we can see in the figure.
For $h \geq h_C = 9$, the ground state in ferromagnetic, and we find
$\Delta S_A = \Delta S_B =0$ as a result of the zero quantum fluctuation in this case as we
discussed in the previous subsection.

We now turn to discuss the XY limit of the model with $w=0$. At $h=0$,
$\Delta S_A = \Delta S_B \approx 0.05$, which is much smaller than the value
in the Heisenberg model ($\approx 0.26$), but is comparable to the value of $0.06$ reported
in the square lattice
XY model~\cite{mit}. As $h$ increases from $0$, $\Delta S_A$ decreases to zero
while $\Delta S_B$ increases to reach a maximum at $h = 3$.  Figs 3 and 4 also show that
$\Delta S_L$ is symmetric with respect to the value of $h =3$ in the region $0 \leq h \leq 6$.
In the XY limit, the quantum fluctuation arising from the
$y$-component of spin is absent. The off-diagonal matrices $M_{\pm}$ in $H_{FL}$ are given by
for $h \leq h_C$,
\begin{eqnarray}
 M_{\pm} = \left ( \begin{array}{ccc}
 0   & \frac{h-3}{6}\gamma_{\vec k} & \frac{h-3}{6} \gamma_{\vec k}^{*}   \\
 \frac{h-3}{6}\gamma_{\vec k}^{*}  & 0 & [\frac{(h-3)^2}{18} - 1 ] \gamma_{\vec k}       \\
 \frac{h-3}{6}\gamma_{\vec k} & [ \frac{(h-3)^2}{18} -1 ]\gamma_{\vec k}^{*}   & 0
        \end{array}
 \right )
\end{eqnarray}
At $h=3$, there is no finite matrix element between the sublattice $A$
and sublattices $B$ or $C$.  Therefore, the triangle lattice is decomposed into a honeycomb
lattice consisting of sublattices
$B$ and $C$  and $N$ isolated spins of sublattice $A$.
Consequently, the spins in $A$ are all parallel to the field with $\Delta S_A =0$,
and $\Delta S_B$ is the same as the result in a honeycomb lattice in the absence of
external fields. The form of $M_{\pm}$ also indicate a symmetry with resepct to
$h=3$, although the rigorous mathematics involves more
delicated analyses since the matrices contain elements with both
even and odd functions of $h-3$.

As shown clearly in the figures, there are sharp peaks for both $\Delta S_A$ and
$\Delta S_B$ at $h = h_c$ except for $w=1$. The peak height increases as $w$ decreases,
and is the largest for $w=0$. Part of this feature found in our
calculations may be understood by examining the ferromagnetic phase for $h \geq h_C$.
In this phase, we have $\Delta S_A = \Delta S_B = \Delta S$, which is given by
\begin{eqnarray}
\Delta S & = & -\frac{1}{2} + \frac{1}{6N} \sum \limits_{n,\vec k}
\frac{-3 + \frac{h}{2} + 3(1+w)T_{n,\vec k}}{\omega_{n,\vec k}}
\end{eqnarray}
From this expression, we see that $\Delta S$ decreases as $h$ further increases as we
can see from the figures.
$(\partial \Delta S / \partial h )_{\mid h \rightarrow h_C + 0^+} \rightarrow - \infty$,
so that the slopes of $\Delta S$ divergent. This explains the singularities of
$\Delta S$ found in our numerical calculation shown in the figures.

\begin{figure}
\includegraphics*[width=3.in,angle=270]{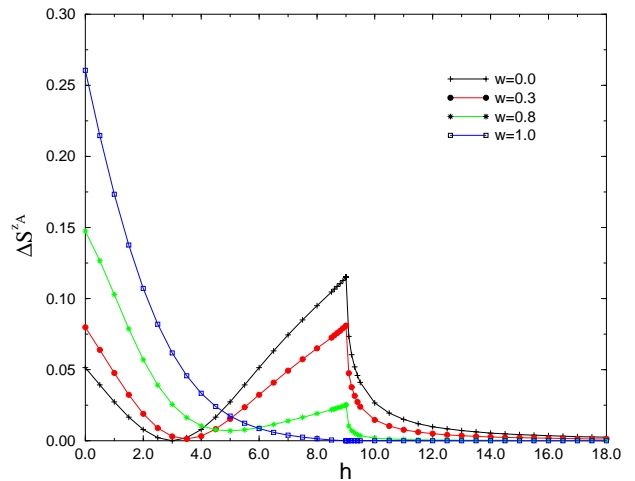}
\caption{Quantum correction to local spin in sublattice $A$ as function of magnetic field
$h$ for several $w$. }
\label{fig:pd}
\end{figure}

\begin{figure}
\includegraphics*[width=3.in,angle=270]{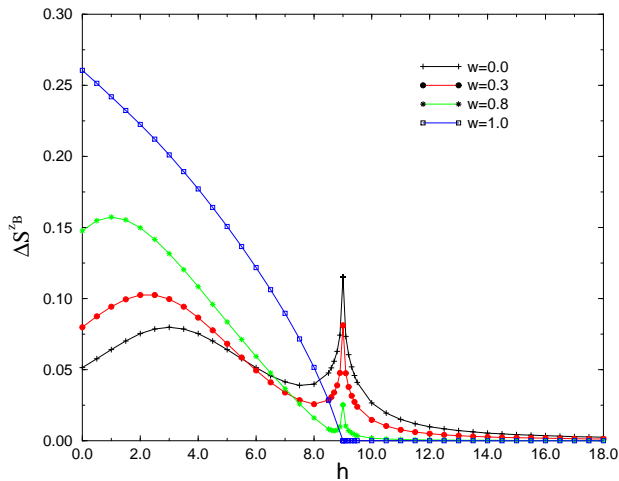}
\caption{Same as in Fig. 3 in sublattice $B$.}
\label{fig:pd}
\end{figure}

The most interesting result we can learn from these calculations
is the possible field enhanced quantum fluctuations in certain quantum
spin systems.  This is somewhat in contrary to the general intuition since
 the field is expected to partially
polarize the spins hence to suppress the quantum fluctuation. Our calculations and
analyses demonstrate that this intuitive argument may break down
in certain parameter region.
This finding should have important implications for the magnetic field driven
quantum phase transition from a spin solid to a spin liquid.
In the spin wave theory, $\Delta S$ is a measure of the quantum fluctuation.
If $\Delta S = S$, the sublatitce magnetization vanishes, indicating the melt-down
of a quantum solid.

\section*{\bf 5. Discussions and Summary}
We have applied the spin wave theory to study quantum XXZ model on a triangle lattice
in the presence of an in-plane magnetic field. Our  model includes the Heisenberg model
at one limit and the XY model at the other limit. We have calculated the ground state
energy and the sublattice magnetization as functions of the external field and 
the anisotropy of the spin coupling. We have found that the field may enhance or suppress
the quantum fluctuation and the reduction of the sublattice magnetization, depending on 
the spin anisotropy and the field strength. At the 
intermediate field and in the  anisotropic case, the field enhances the quantum fluctuation.
The reduction of the sublattice magnetization in the model we have studied 
is still too small to
destroy the spin long range order even for the smallest spin system with $S=1/2$. 
This is because the 
model we studied here is still quite far from the solid-liquid 
boundary in the absence of the field.  Nevertheless,
the qualitative effect of the field on the quantum fluctuation we have found 
is interesting, and it may have
important implications for systems near the solid-liquid phase boundary. 
Near that boundary, the external magnetic field may well drive a quantum solid 
to a quantum liquid due to the increase of the quantum fluctuation. 
In quasi 2-dimensional triangle lattice of $Cs_2CuCl_4$, the in-plane spin-spin couplings
are strongly anisotropic, and the coupling along one 
direction is about three times as large as those along the other two 
directions. The spin system is between one and two dimensions, and the system in the absence of field
has a long range order but is closer to the 
spin liquid boundary. In that type of systems, an enhanced quantum fluctuation 
may easily destroy the long range order. 
We speculate that the in-plane field induced 
phase transition~\cite{coldea3} to the quantum liquid state observed in $Cs_2CuCl_4$ could be
due to the increase of the quantum fluctuation. More detailed calculations on
more realistic model would be interesting to carry out to see if such a conjecture 
is relevant to the experiment.

We would like to thank D. A. Tennant, R. Coldea for discussions on their experiments.
We also wish to thank S. Q. Shen for many useful discussions on this subject, 
especially in the early stage of the work.  
This work is in part supported by Chinese Academy of Sicences,
and by the US DOE grant DE/FG03-01EG45687.

\end{document}